\documentstyle[12pt]{article}
\def\b{\bigskip}
\def\m{\medskip}
\def\s{\smallskip}

\def\ki2{$\chi^2$}
\def\is{interstellar}
\def\ism{interstellar medium}
\def\kms{km.s$^{-1}$}
\def\cm{cm$^{-2}$}

\def\caii{Ca{\sc ii}}
\def\nai{Na{\sc i}}
\def\cii{C{\sc ii}}
\def\civ{C{\sc iv}}
\def\ovi{O{\sc vi}}

\def\siiii{Si{\sc iii}}

\def\feii{Fe{\sc ii}}
\def\mgii{Mg{\sc ii}}
\def\mgi{Mg{\sc i}}
\def\hei{\mbox{He\,{\sc i}}}
\def\heii{\mbox{He\,{\sc ii}}}
\def\hi{\mbox{H\,{\sc i}}}
\def\di{\mbox{D\,{\sc i}}}
\def\oi{\mbox{O\,{\sc i}}}
\def\ni{\mbox{N\,{\sc i}}}
\def\dsh{D{\sc i}/H{\sc i}}
\def\lya{Lyman $\alpha$}

\begin{document}

\vskip 1.6truecm
\begin{center}
{\bf The Local Interstellar Medium}\\
\b\b\b
{Roger Ferlet}\\
\m
{Institut d'Astrophysique de Paris, CNRS}\\
{98 bis boulevard Arago, 75014 Paris, France}\\
{e-mail: ferlet@iap.fr}\\
\end{center}
\vskip 3truecm

SUMMARY
\baselineskip 0.5cm
\bigskip
\s

Substantial progress in the field of the Local Interstellar Medium has been 
largely due to recent launches of space missions, mostly in the UV and X--ray
domains, but also to ground-based observations, mainly in high resolution
spectroscopy. However, a clear gap seems to remain between the wealth of new
data and the theoretical understanding. This paper gives an overview of some
observational aspects, with no attempt of completeness or doing justice to all
the people involved in the field. As progress rarely evolves in straight paths,
we can expect that our present picture of the solar system surroundings is not
definitive.

\b\b

KEY WORDS

\m
\noindent
Galaxy (the): solar neighbourhood (07.32.1)\\
Interstellar medium: clouds: individual: the Local Cloud (09.11.1)\\
Interstellar medium: bubbles (09.08.1)\\
Interstellar medium: kinematics (09.17.1)\\
Interplanetary medium (09.06.1)

\b\b
INTRODUCTION
\baselineskip 0.7cm
\bigskip
\smallskip

The Local Interstellar Medium (LISM) has been an active area of investigation 
since the very beginning of the seventies, but for quite long it was uncommon 
to appreciate what was being learned in this area by others. The story began 
as many widely dispersed rivulets which converged for an IAU Colloquium on 
Madison, Wisconsin in June 1984 dedicated to the LISM (Kondo et al. 1984). 
Since then, several LISM meetings took place, the last one (again IAU) in April 
1997 in Garching, Germany (Breitschwerdt et al. 1998).

There is no exact definition of the LISM, neither in terms of column densities
nor in terms of distances, since it comprises a number of components at very
different densities and temperatures distributed in highly asymetric ways. In
fact, the study of the LISM is crucial because it offers two unique 
opportunities. First, the physical state and the multi-phase structure of the 
LISM bears important consequences for our picture of the more general ISM. Very 
little is yet known on how do coexist the hot, warm and cold phases, i.e. on 
the filling factors, the characteristics of the interfaces, the pressure balance
between phases, the ionization degrees. Thanks to the low column densities
encountered in the LISM, it should be possible to disentangle individual clouds
(if they exist) and understand their subtle properties and relations. Obviously
also, all the wavelengths reach our telescopes after passing through the LISM.

Second, we want to know about our environment. The physical state in the 
immediate \is~vicinity of the Sun represents the boundary conditions for the 
expansion of the solar wind and constrains its confinement, i.e. the size of 
our heliosphere. While galactic tides equal the gravitational attraction of the 
Sun at distances of $\sim$~150 000 AU in the galactic plane (the nearest 
star $\alpha$~Cen being at about 3$\times$10$^5$~AU), it is surprising that the 
much closer location of the heliopause is still uncertain. More generally, the
interpretation of heliospheric observations critically depends on the physical
conditions that ions, dust grains and cosmic rays encounter on their way to the
solar system.

Last but not least, the galactic environment of the Sun influences the one AU
interplanetary volume around the Earth, including the Sun--Earth coupling
mechanisms and the high energy radiation and particle fluxes at the top of the
Earth's atmosphere. The Sun moves through space (with respect to the Local
Standard of Rest, LSR, defined by the velocity ellipsoid of nearby cool stars
which have relaxed to dynamical equilibrium in the galactic gravitational field,
Mihalas and Binney 1981) with a velocity of 16.5 pc per million years and 
encounters between the Sun and \is~clouds do occur. Observations of "dense" 
\is~structures down to AU scales (Frail et al. 1994) show that environment 
variations are even possible on time scales down to years. Understanding the 
distribution, kinematics and properties of nearby \is~clouds allows to 
evaluate such encounters which may perturb the terrestrial climate and 
eventually imprint anomalies in the geologic records. Extreme changes due to 
"galactic weather" might not favor the emergence of life.

\b\b
GLOBAL MORPHOLOGY AND DISTRIBUTION
\b\s

Within 500 pc of the Sun and 50 pc of the galactic plane, not less than about
15 molecular clouds are seen through the CO molecule which is a standard tracer 
for densities 10$^3$--10$^6$~cm$^{-3}$~(Dame et al. 1987). They define the 
positions of spiral arms and the local warp of the galactic plane associated 
with the Gould's Belt. They roughly represent about 50\% of the mass of 
\is~matter and their filling factor is estimated to be 0.16. They are basically
at rest in the LSR. There is an overall pronounced deficiency of \is~matter in 
the third Galactic quadrant between l=220$^o$~and l=250$^o$~which corresponds 
to the interarm region between the Local and Orion arms and merges with the 
interarm region between the "grand design" Orion and Sagittarius spiral arms 
at about 1 kpc from the Sun. This deficiency is also shown by a lack of diffuse 
gamma-ray emission (Hunter et al. 1997) and of 21 cm emission (Stacy and 
Jackson, 1982). These last authors even identify an optically thin \hi~window 
out to the edge of the Galaxy near l=245$\pm$5$^o$~and b=$-$3$\pm$3$^o$. CO has
also revealed a population of high-galactic latitude ($\mid$b$\mid\geq$25$^o$) 
molecular clouds with a mean distance of $\sim$100 pc (Magnani et al. 1985). 
One of these, MBM 12 at l=159$^o$, b=$-$34$^o$, has been proved to be located 
at $\sim$65 pc (Hobbs et al. 1986), making it the nearest known molecular cloud.
There is another gap in the distribution of this population in the direction 
180$^o$<l<340$^o$~which mimics the above one.

The distribution of diffuse dust and gas of densities 1--1000 cm$^{-3}$~can be 
traced by observing the reddening of starlight by \is~dust (Lucke 1978). The 
minimum contour E(B$-$V)=0.1, which corresponds to a hydrogen column density
of $\sim$5.8$\times$10$^{20}$~\cm, follows as expected the spiral arm molecular
material. An abrupt increase in reddening at 200 pc is seen in directions
l=258.4$^o$, b=$-$11.1$^o$~l=271.3$^o$, b=+5.3$^o$, which is attributed to the
onset of reddening due to dust in the Gum nebula (Franco 1990). The Vela sheet
(l=272--279$^o$, b=$-$3$^o$) is embedded in the Gum nebula.

Interstellar absorption lines spectroscopy against stellar continua provide 
perhaps the most powerful tracer of diffuse clouds. For short path lengths
column densities are small and resonance lines are the most opaque lines for 
each ion. Optical neutral sodium D lines and singly ionized calcium K lines 
are easily observed and sample densities 0.1 to 100 cm$^{-3}$~and thermal 
temperatures 50 to 8000 K. But most of the resonance lines show up in the UV 
domain and allow to sample virtually all nearby \is~media.

Already at the time of the Madison meeting, two characteristics of the \is~
matter distribution within $\sim$200 pc were recognized thanks to absorption
studies: inhomogeneity and asymmetry. Plots of column densities versus distances
clearly show that the column density is not a strictly monotonic function of 
distance (see e.g. Frisch and York 1991). Early maps of contours of equal 
hydrogen column density (e.g. Frisch and York 1983; Paresce 1984) showed a 
noticeable rapid increase towards the Sco--Oph direction, more or less towards 
the Galactic center. Although these authors grossly disagreed about the 
distance to this increase, they however both also realized the lack of neutral 
gas towards the third quadrant, in particular in the Canis Majoris direction 
(l$\sim$240$^o$) which was subsequently identified as a true "tunnel" nearly
free of matter (Gry et al. 1985, 1995; Cassinelli et al. 1996).

In these early maps, N(\hi) was directly derived from \lya~observations with 
the resolving power R=$\lambda/\Delta\lambda\sim$20 000 of the Copernicus
satellite ($\sim$15 \kms) or even a factor of two lower of IUE, or indirectly 
infered from other ions sensitive to both ionized and neutral material. 
Thereafter, \nai~was sampled by Welsh et al. (1990, 1991) and \caii~by 
Vallerga et al. (1993a) towards 45 early-type stars within 220 pc from the Sun 
with a resolving power R=$\sim$150 000. More specifically, the Sco--Cen 
association was extensively studied by Crawford (1991), in both \nai~and 
\caii~but at a slightly lower resolution, and Loop I near l=320$^o$,
b=20$^o$ by Centurion and Vladilo (1991) and Fruscione et al. (1994). The 
overall distribution was confirmed, in particular: the abrupt increase in 
column densities towards Sco--Oph beyond roughly 130 pc, a N(\hi) rise near the
distance of about 70 pc and the CMa tunnel. 

For the over 80 absorption components detected in \caii~(Vallerga et al. 1993a),
the mean LSR velocity is 0.9 \kms~with an rms of 11.3 \kms. Identical results
apply to \nai, but with a smaller rms of 3.6 \kms. The statistical analysis of 
a set of northern hemisphere neutral hydrogen clouds observed at 21 cm gives
also an essentially zero LSR velocity with a dispersion of 6.9 \kms~(Belfort and
Crovisier 1984). These clouds are therefore, in a global sense, associated with
the general motion of the solar neighborhood and not of the Sun itself. 
Specifically from the Sco--Cen association, a general outflow is confirmed 
(Crawford 1991).

Moreover, when plotted against the LSR velocities, the N(\nai)/N(\caii) ratios 
for each identified absorption feature show the famous Routly--Spitzer (1952) 
effect in which higher velocity clouds have lower ratios. Since calcium is
generally highly depleted in the \ism, this effect is widely interpreted as due
to a mechanism of desorption/destruction of grains efficient at returning 
calcium to the gas phase. According to Vallerga et al. (1993a), the effect 
seems evident down to $\pm$10 \kms, which supports a mechanism efficient even at 
these low velocities.

\b\b
THE LOCAL BUBBLE
\b\s

In the early 70's, the brightness distributions of soft X-rays (below 2 keV)
were available for most of the sky. Due to photoelectric absorption by neutral
hydrogen, their mean free path becomes less than the radius of the Galaxy. Thus,
for the ultrasoft component (0.08--0.3 keV) a local galactic origin seemed 
compelling. Meanwhile, a widespread hot phase of the \ism~has been recognized 
from ubiquitous \ovi~absorption lines observed with the Copernicus satellite
(Jenkins and Meloy 1974). A link between the soft X-ray emission and a supernova
origin rapidly became suggestive (McKee and Ostriker 1977). However, a hot 
medium with the typical temperature of $\sim$5$\times$10$^5$~K of the McKee 
and Ostriker's model fails to reproduce the ratios between the C (160--284 eV), 
B (120--188 eV) and Be (77--111 eV) bands observed with the WISCONSIN rocket
survey (see e.g. McCammon and Sanders 1990). This famous model is now thought 
to be irrelevant to further useful discussion (Cox 1995).

An obvious interpretation of the observed ultrasoft background is the existence
of a local cavity, filled with hot plasma and devoid of neutral hydrogen.
Assuming that {\it all} of this background originates in the cavity, it was
thought that the measured X-ray intensity along a given line of sight was
directly proportional to its extent. Thus, a Local Hot Bubble was conceived,
extending about 200 pc perpendicular to and very much less into the galactic 
plane, with a complicated 3-dimensional shape (Snowden et al. 1990) and a
temperature of 10$^6$~K obtained from a collisional ionization equilibrium 
model.

In the M--band (0.5--1.0 keV) which samples more distant soft X-ray emission 
than the B--band, the WISCONSIN survey, later confirmed by the ROSAT All Sky 
Survey, found the sky brighness away from some identified prominent sources 
(Loop I, Cygnus superbubble, Eridanus cavity etc.) to be fairly isotropic. 
This is not easily explained because discrete sources would certainly exhibit 
intensity variations with latitude. In fact, this has marked the begining of a 
fundamental change in our concept of the Local Bubble thanks to space missions 
like EUVE, DXS and ROSAT.

One of the first deep pointed ROSAT observations were the so-called shadowing
experiments. Snowden et al. (1991) found that the X-ray intensity of a line of
sight passing through the Draco nebula is substantially attenuated, i.e. that a
minimum of 50\% of the C--band emission is from beyond the cloud. With distance
limits between 300--1500 pc, at least half the flux was clearly from
{\it outside} the Local Hot Bubble, in contradiction with the standard 
assumption that all the soft X-ray background originates in the cavity. For the
nearby heavy absorber MBM 12, almost no C--band shadow was observed while less 
than 30\% of the M--bands emission is foreground (Snowden et al. 1993). ROSAT 
has undoubtedly established the \is~origin of the soft X-ray background but also
the existence of such a diffuse emission from beyond the "previous" Local 
Bubble.

The Diffuse X-ray Spectrometer (DXS) experiment was an attached 1993 Shuttle 
payload consisting of two Bragg crystal spectrometers with "good" resolution
over the range 0.15--0.284 keV (C--band, 8.3--4.4 nm). Emission lines and 
blends were detected, thus confirming the thermal origin of the low latitude 
diffuse background: a hot phase of the \ism. However, no collisional ionization 
equilibrium model (with cosmic abundances or depleted) at any temperature in 
the 10$^5$--10$^7$~K range can adequately reproduce the observed spectra 
(Sanders et al. 1996). The widths of the \ovi~absorption lines discovered by 
Copernicus can only provide an upper limit of the temperature, thus inconclusive
for deciding in favor of cooler non-equilibrium models. A hot plasma model 
can further be dismissed by the observation of two lines of sight crossing 
the Local Bubble: i) towards $\beta$~CMa ($\approx$200 pc) which shows an 
average electron density <n$_e> \sim$2$\times$10$^{-2}$~cm$^{-3}$~and a very 
low temperature $\leq$5$\times$10$^4$~K (Gry et al. 1985), and ii) towards the
pulsar P0950+08 ($\approx$130 pc) whose dispersion measure gives
<n$_e>$~$\sim$2.3$\times$10$^{-2}$~cm$^{-3}$~(Reynolds 1990). Nevertheless, i)
might not be typical and ii) might be already beyond the Local Bubble.

Due to the short mean free path of EUV photons, shadows are much deeper than for
soft X-rays; they are therefore an excellent probe of the very local \ism. Most
of the flux from a plasma in collisional ionization equilibrium around 10$^6$~K
should appear as emission lines in the EUV. The only lines detected in the
16--74 nm range of the EUVE satellite were \hei~and \heii, with intensities
consistent with local geocoronal and/or interplanetary scattering of solar 
radiation (Jelinsky et al. 1995). These authors derived upper limits for the 
plasma emission measure that are a factor of 5 to 10 below what is expected in 
the B-- and C--bands from the equilibrium plasma model with solar abundances 
over the temperature range 10$^{5.7}$--10$^{6.4}$~K. Exploring possible 
scenarios that could reconcile this discrepancy, the authors favored the 
recent (less than 10$^5$--10$^6$~years ago) heating of the hot gas responsible 
for the diffuse soft X-ray background by an active blast wave, most likely 
caused by a supernova. The dominant features at T$\sim$10$^6$~K are high 
ionization lines of iron and silicon, two refractory elements normally strongly
depleted onto dust grains in the cold \ism. Since evaporative mechanisms that 
return them to the gas phase (e.g. thermal sputtering in shocks) and then raise 
them to high ionization states require timescales longer than 10$^5$~years (Cox 
and Reynolds 1987), the diffuse soft X-ray emitting gas may have depleted 
abundances which would change the spectral distribution and explain the 
non-detection of Jelinsky et al. (1995). However, such models can only be 
constrained by real detections of emission lines, not upper limits. The DXS 
experiment has shown that the data could not be successfully fit with depletion.

\b\b
SMALL SCALE MORPHOLOGY
\b\s

As early as 1978, Vidal--Madjar and collaborators suggested for the first time
that a small \is~cloud might be present very close to the solar system. On the
basis of the observed gradients in both the hydrogen density and the far UV
\is~flux, and of the possible variations of the local \dsh~ratio, they inferred
a distance of few hundredths of a parsec towards the Galactic center. From the
polarization of starlight within 35 pc of the Sun, Tinbergen (1982) claimed the
existence of a patch of \is~material corresponding to E(B$-$V)$\sim$0.003 or
N(H)$\sim$1.5$\times$10$^{19}$~\cm, also in the general direction of the 
Galactic center, with the nearest weakly polarized star beeing at 5 pc. The 
simple picture which emerged from the Madison Colloquium located the Sun at the 
inner edge of a cloud -- the Local Interstellar Cloud (LIC) -- believed to be 
a single entity immersed within a large, very hot bubble (the Local Bubble).

Absorption spectroscopy is the only method able to probe the nearest \is~gas
outside the solar system, provided the spectral resolution and signal to noise 
(S/N) ratios are high enough. Separations between adjacent velocity components
(clouds or gas moving with the same bulk velocity) in the range 1--2 \kms~have 
already been identified (Crawford and Dunkin 1995; Welty 1997), requiring thus 
R$\geq$~few times 10$^5$. \hi~column densities of a few 
$\times$10$^{17}$~cm$^{-2}$~typical for short sight lines translate into 
equivalent widths in the 10$^{-4}$~nm range for common ions in the visible, 
implying S/N ratios well above 100 to be detected. Many UV lines are much 
stronger, but the relatively low resolution and S/N of IUE data only allowed to 
confirm the pervasive presence of low densities towards the North Galactic Pole 
and anticenter region (see e.g. the survey of \mgii~lines in 51 cool stars 
within 30 pc by Genova et al. 1990). With the launch of the Hubble Space 
Telescope, the situation greatly improved. But even with the 3.6 \kms~resolution
of the GHRS, it is estimated that only 10\% of the \is~velocity components are 
resolved for unsaturated lines (Welty et al. 1996).

The complexity of the local gas has been definitively demonstrated when Ferlet 
et al. (1986), using the R=10$^5$~Coud\'e Echelle Spectrometer at ESO--La Silla,
detected three velocity components with a total equivalent width of 0.017 nm
in the \is~\caii~spectrum of $\alpha$~Aql (Altair) at 5 pc from the Sun. 
Subsequent \caii~surveys (Lallement et al. 1986; Lallement et al. 1990; 
Lallement and Bertin 1992) further confirmed this complexity. Most of the stars 
within 20 pc exhibit two or three absorption components, showing that the solar 
vicinity comprises different cloudlets. Similarly, Vallerga et al. (1993a) 
detected several \caii~components to most of their stars closer than 50 pc 
except in the third Galactic quadrant. However, Welsh et al. (1990, 1991) 
failed to detect \nai~in the same sample at less than 42 pc, their S/N ratio
corresponding to N(\hi)<2.5$\times$10$^{18}$~\cm. Studying the spatial 
correlation of the \caii~velocities within 50 pc, Vallerga et al. (1993a) also
showed that in fact most components do not have a counterpart (at a precision 
of 2 \kms) in a nearby sightline. They concluded that most \caii~clouds subtend 
angles less than 15$^o$. With an assumed mean distance of 30 pc, this 
corresponds to a typical upper limit of 7 pc for the size of these clouds.

The three components in $\alpha$~Aql have also been seen in the \mgii~and \feii~
GHRS data (Lallement et al. 1995). Towards the same region, pronounced angular 
variation in N(\caii) within $\sim$20 pc are known: log N(\caii)=11.53 \cm~ 
towards $\alpha$~Oph (Crawford and Dunkin 1995) whereas towards $\lambda$~Aql 
log N(\caii)<9.59 \cm~(Vallerga et al. 1993a). Towards Vega at 7.5 pc, also 
three components are identified in \feii~(Lallement et al. 1995). Even towards 
Sirius ($\alpha$~CMa) at 2.7 pc, two absorption features are detected in 
\mgii~and \feii, with log N(\hi)=17.23 \cm~(Lallement et al. 1994; Bertin et 
al. 1995). Indeed, the volume of the Local Bubble could contain about 2000 
"standard diffuse clouds" such as those statistically explored via extinction 
studies (Spitzer 1985) and assumed to be spherical. An even larger number could
be expected according to the considerably more complete description provided 
by optical absorption-line and 21 cm studies.

\b\b
KINEMATICS
\b\s

A major issue in the local \ism~studies has long been the identification of a
velocity vector that would show a coherent motion of the local medium. In that
respect, ground-based data are most useful because of their resolution and their
excellent accuracy of absolute wavelength scale. With a simple Doppler
triangulation, it is possible to reconstruct a velocity vector from several (at
least three) measured radial velocities, provided the cloud is close enough to 
the Sun. Following the first attempt by Crutcher (1982) who used only seven 
lines of sight shorter than 100 pc, several different vectors were found.
Although in disagreement with each others, all of them indicated nevertheless
that material flows more or less from the higher density regions towards the 
empty region. Furthermore, they were also in disagreement with the velocity of 
the \is~flow in the inner solar system.

The Sun is moving through the tenuous gas of the LIC, thus creating a flow of
neutral hydrogen and helium atoms from \is~origin which enter the solar system
(Fig. 1). These atoms are detected through resonance scattering of solar photons
by in-situ experiments. Based on two independent determinations -- the H 
\lya~and the He 58.4 nm glows -- the direction of the \is~wind within the inner 
interplanetary medium was found to be towards: l=184.3$\pm$3.5$^o$~and 
b=$-$15.8$\pm$3.4$^o$~(Bertaux et al. 1985). It is in very good agreement with 
the direct detection of the \is~\hei~with the Neutral Gas Experiment onboard 
the Ulysses spacecraft (Witte et al. 1993). However, the velocity modulus has 
long been a matter of discussion. Hydrogen observations give a modulus of 20 
\kms, whereas 26 \kms~is deduced from helium (Bertaux et al. 1985; Witte et al. 
1993).

Using the Doppler triangulation with visible high resolution spectroscopic 
observations at high signal to noise ratio of very nearby stars performed at 
both ESO and Observatoire de Haute Provence, Bertin et al. (1993a) identified 
the LIC flowing from l=6$^o$~and b=+16$^o$, with an uncertainty of $\pm$3$^o$,
and measured a velocity modulus of 25.7$\pm$1 \kms~with respect to the Sun. 
These are precisely the characteristics derived through helium backscattered 
data. They were further strongly confirmed with HST data towards Capella 
(Linsky et al. 1995) and several other stars (Lallement et al. 1995; Bertin et 
al. 1995). Together with the Ulysses result, this brings a definitive proof of 
the LIC identification. An immediate important consequence is the existence of 
a perturbation, namely a deceleration of $\sim$6 \kms, of neutral hydrogen 
atoms when they travel through the heliospheric interface. While the helium 
initial velocity of 26 \kms~is thought to be unaffected because of the weak
collisional cross-section of helium with solar protons, hydrogen atoms could on
the contrary be primarily coupled to the plasma via charge--exchange reactions 
and thus expected to be decelerated and heated beyond the heliopause, as 
predicted by heliospheric models (Baranov et al. 1991; Osterbart and Fahr 1992; 
Qu\'emerais et al. 1992).

\b\b
THE LOCAL INTERSTELLAR CLOUD
\b\s

1 Structure
\b

In order to construct the structure of the cloud surrounding the solar system, 
one may use the column density of the components detected at the projected LIC 
velocity towards nearby stars. Whereas the total integrated N(\hi) is well 
determined through the \lya~line, it is not possible to derive the velocity 
structure from this line alone because of its heavy saturation even for 
N(\hi)$\sim$10$^{17}$ cm$^{-2}$. It is therefore necessary to observe resonance 
lines from other elements at the highest possible resolution, usually \mgii~and 
\feii. Introducing then the detected velocity structure, one may infer N(\hi) 
for individual components. However, although both \mgii~and \feii~are the 
dominant ionization stages in \hi~regions, they have the drawback of also 
sampling some ionized gas as their ionization potentials are slightly larger 
than for \hi. Furthermore, in the case of late-type stars, the closest ones,
\is~\lya~is superposed on their chromospheric emission and the modelling of 
that "continuum" can induce potentially large uncertainties. This type of study
has been done with HST for more than a dozen stars closer than 20 pc. The LIC 
is identified in all of them but $\alpha$~Cen (Lallement et al. 1995; Linsky et 
al. 1995; Ferlet et al. 1995; Linsky and Wood 1996; Piskunov et al. 1997; Wood
and Linsky 1998).

Assuming that the volume density is constant in the LIC, column densities 
draw its contours. The LIC is flattened in the galactic plane and extends mainly
in the first quadrant (towards Altair, Vega). The Sun is located at its very 
edge towards $\alpha$~Cen (less than about 30000 AU) and at less than 1 pc from 
its surface in the direction of Canis Majoris (Fig. 2).

It has to be recalled that at 1.3 pc, $\alpha$~Cen is the closest star inspected
for absorption lines, whereas the distance to the heliopause (still not 
precisely known) is only about 0.0005 pc. Therefore, the inferred averaged
physical cloud properties might not necessarily correspond exactly to the
closest solar system surroundings. For instance, a dense 
($\sim$10$^5$ cm$^{-3}$), cold ($\sim$50 K), very thin ($\sim$5--10 AU) cloud
with a typical diffuse \is~depletion would be barely detectable, especially if
blended with adjacent higher column density and/or warmer gas. These kinds of 
uncertainties might explain why the LIC is not detected towards $\alpha$~Cen
(Lallement et al. 1995): it could be hidden by the six times larger absorption
due to the main component in that sight-line.

The other nearby cloud seen towards $\alpha$~Cen (often called the G cloud for 
Galactic center) is of course at less than 1.3 pc from the Sun. This G cloud is
also possibly seen towards Altair but for this sight line it is blended with 
the LIC. Apart from the G cloud and the LIC, six additional components are 
detected towards Sirius, Procyon, Altair, $\alpha$~PsA and Vega, i.e. at less 
than 7.5 pc from the Sun (Fig. 2). If one believes that all these components are
indeed distinct cloudlets outside the LIC, the highest column density sight 
lines in the first quadrant imply a density larger than $\sim$0.2 cm$^{-3}$~for 
the LIC (see below). Furthermore, the two non-LIC clouds seen towards Sirius 
(2.7 pc) and Procyon (3.5 pc) 25$^o$~apart in the sky are different which 
implies for these clouds dimensions of about 1 pc or smaller and locations at 
or just beyond the edge of the LIC.
\b\s

2 Motion
\b

The excellent agreement between the observed Doppler shifts and those predicted
by projecting the LIC/interplanetary wind velocity vector onto target directions
covering a very large fraction of the whole celestial sphere (Lallement et al. 
1995) secures the LIC motion. Furthermore, the solar apex direction is nearly
perpendicular to the LSR upwind direction of the LIC. As a consequence, the 
Sun has entered the LIC tens of thousands years ago at most; it is leaving it 
and will enter the G cloud in less than 70000 years (1.3 pc at 18 \kms). The G 
cloud being more rapid than the LIC, both might be already in contact. If not, 
the solar system would go through the Local Bubble. However, one must keep in 
mind the possibility to interpret the observed velocity components not as due 
to physically identified clouds but as the manifestation of large bulk motions 
like mesoturbulence or as due to \is~shock waves.

In the LSR velocity frame, the LIC is moving towards us from the direction 
l$\sim$335$^o$, b$\sim-$2$^o$. From a compilation of the velocities of all 
\is~components observed for stars within about 100 pc of the Sun, Bertin (1994) 
has shown an overall motion of the \is~gas from about the same direction (the 
Sco--Cen association), again after correcting from the solar proper motion, 
which is not an effect of the galactic rotation. Moreover, a velocity gradient 
seems to be associated with this coherent flow: decreasing velocities towards 
the downwind direction. The dispersion shows, however, that the flow is not 
uniform. The most plausible explanation of such a situation is an acceleration 
through a shock front. The energy needed is largely available in a supernova 
explosion.

For all absorbing gas within 30 pc of the Sun in which N(\nai)/N(\caii) has 
been measured, this ratio is smaller than 0.5 and even smaller than 0.2 for
material (including the LIC) within 15 pc (Bertin et al. 1993b). In dense cold 
clouds, this value is much higher (10 to 100), refractory elements such as 
calcium being heavily depleted onto dust grains. Since the local clouds are 
moving at high velocities, in the context of the Routly--Spitzer effect these 
low ratios are strong evidence of calcium returned to the gas phase through 
destruction of grains by an \is~shock. Similar anomalous abundance patterns 
(lower depletions) are found towards $\alpha$~Oph (Frisch et al. 1987), 
$\eta$~UMa (Frisch and York 1991), $\alpha$~CMa (Bertin 1994), $\alpha$~Aur and 
$\alpha$~CMi (Linsky et al. 1995) and $\alpha$~Cen (Linsky and Wood 1996). 
These last authors have even shown that \is~metal abundances (measured through
magnesium) do vary significantly between the LIC and the G cloud. Piskunov et 
al. (1997) also found large variations over distances of only a few parsecs
within the LISM.
\b\s

3 Density, pressure and ionization
\b

Thanks mainly to EUVE observations of the continua and photoionization edges
of \is~\hei (50.4 nm) and \heii~(22.8 nm) towards a large set of white dwarfs,
values of the \hi~density in the LISM are constrained to lie between 0.15 and
0.34 cm$^{-3}$~(Vallerga 1996). This is consistent with analyses inside the 
heliosphere which give 0.165$\pm$0.035 cm$^{-3}$~(Qu\'emerais et al. 1994),
although some filtration of the \hi~through the heliosphere interface might be
implied. Using this \hi~density, one deduces a thermal pressure (P/k=nT)
$\sim$1200 cm$^{-3}$~K, a factor of about 10 lower than the Local Bubble
pressure (P/k$\sim$10000--20000, Bowyer et al. 1995). Already recognized at the 
time of the Madison meeting, that discrepancy has fuelled a controversy over
competing models of the galactic ISM (Cox 1995). This is an unstable situation
that requires other pressure sources such as magnetic fields if one wants to 
maintain a pressure confined local cloud.

Also from EUVE observations, helium is on average 25\% ionized and even seems 
to be systematically more ionized than hydrogen (n$_{\hi}$/n$_{\hei}\approx$14,
Dupuis et al. 1995; Barstow et al. 1997), which is counterintuitive considering 
the much larger ionization potential of \hei~relative to \hi. The most likely 
explanation for Jelinsky et al. (1995) is a recent ($\leq$10$^6$~yrs) ionizing 
event. However, \is~parameters derived from EUVE observations depend somewhat 
on white dwarf model atmospheres. More importantly, the limited spectral
resolution of EUVE prevents to resolve individual kinematic components and
therefore only total column averages are available, the LIC being hidden in 
them.

The ionization state of the nearby gas is a long standing problem which had to 
be re-examinated when Vallerga et al. (1993b) discovered with EUVE that the B2
II star $\epsilon$~CMa is 30 times brighter at 60 nm than HZ43, the brightest 
white dwarf in the sky. One of the more precise methods to derive the electron 
density n$_e$~is to measure the different ionization states of the same element.
\mgi~and \mgii~are widely used, but prior to the high resolution of GHRS onboard
HST, only average values were obtainable. Although \mgi~is very faint in the 
LIC, it has been unambiguously detected. Provided equilibrium conditions 
prevail, the measured abundance ratios lead to 
n$_e$=0.09$^{+0.23}_{-0.07}$ cm$^{-3}$~towards $\epsilon$~CMa (Gry et al. 1995),
n$_e\sim$0.1--0.6 cm$^{-3}$~towards $\delta$~Cas (assuming T=7000 K) and even 
larger values towards Sirius (Lallement and Ferlet 1997). Indeed, at least the 
last value is unrealistically large. As a matter of fact, the equilibrium 
assumption might be somewhat wrong because the hydrogen recombination time in 
the diffuse \ism~can be very long: the distance covered before a recombination 
can be estimated as 30 pc which is probably not negligible relative to the 
spatial scale of the EUV flux variations (known to be strongly anisotropic). 
Furthermore, large errors on n$_e$~are likely to occur in view of the strong 
temperature dependence of the \mgi~ionization rate.

An independent way to evaluate n$_e$~is to use the measured \nai/\caii~ratio. 
But here again, the equilibrium assumption is needed, as well as an estimation 
of either the calcium depletion or the neutral hydrogen density. Lallement and 
Ferlet (1997) derived in that way n$_e$~of the order of 0.05 cm$^{-3}$~towards 
$\delta$~Cas. Although the error bar overlaps with the range derived from 
magnesium, both methods are quite discrepant. Another way which has the 
advantage of being independent of assumptions about the ionization equilibrium 
is to use the ground state and excited state column densities derived from the 
\is~\cii~absorption features. Towards Capella, this method implies 
n$_{e}\sim$0.11 cm$^{-3}$~(Wood and Linsky 1997). However, error bars are still
large. As already noted, parameters such as column densities are subject to 
errors in the location of the continuum which the lines are absorbing. In the 
case of cool stars like Capella, this continuum is in fact the unknown stellar 
emission line.

For all these methods, the knowledge of the gas temperature is essential. From 
profile fitting of \feii~and \mgii~lines towards Sirius, Lallement et al. (1994)
derived a LIC temperature T=7600$\pm$3000 K. From the \is~neutral helium within 
the solar system, Witte et al. (1993) derived T=6700$\pm$1500 K. The Capella 
and Procyon lines of sight (Linsky et al. 1995) have provided perhaps the most 
accurate values for the LIC: T=7000$\pm$900 K together with a most probable 
speed for the assumed gaussian nonthermal motions (turbulence) of 1.6$\pm$0.4 
\kms. If there is an agreement around 7000 K, error bars are still large enough
to prevent a very precise evaluation of the electron density.

A new approach has been developped to provide n$_{e}$~in the Local Cloud. On 
the basis of the invariance principle which relates directly the properties
of two media separated by a layer of material -- the LIC properties as observed
from the Earth orbit and the LIC material outside the heliosphere -- Puyoo et 
al. (1997) derived n$_{e}\sim$0.044 cm$^{-3}$, a value significantly lower than 
most of the previous ones, except towards a pulsar (0.02 cm$^{-3}$, Reynolds 
1990). The same authors also derived n$_{\hi}\sim$0.24 cm$^{-3}$~and 
n$_{\hei}\sim$0.023 cm$^{-3}$. Towards Capella and Procyon, Linsky et al. (1995)
found respectively n$_{\hi}\sim$0.04 cm$^{-3}$~and $\sim$0.1 cm$^{-3}$, but 
both stars lie well outside the LIC. On the contrary, Linsky and Wood (1996) 
derived n$_{\hi}$=0.15 cm$^{-3}$~with an uncertainty of a factor of two towards 
$\alpha$~Cen which lies inside (or almost) the G cloud.

The very low ionization claimed for the LIC by Puyoo et al. (1997) has been
very recently confirmed in the course of detailed analysis of the lines of 
sight towards Sirius B (H\'ebrard et al. 1998) and the nearby white dwarf 
G191--B2B (Vidal--Madjar et al. 1998; see also below). In these sight-lines, 
no \siiii~absorption is detected in the LIC (nor in the other component towards
Sirius), in contrast with two much longer CMa lines of sight (Gry et al. 1985, 
1995). Except if the LIC is extremely inhomogeneous, a very likely explanation 
is a random superposition in the velocity space of the LIC with the very long 
ionized CMa tunnel (1 pc relative to 200 pc!).
\b\s

4 The deuterium abundance
\b

It is widely accepted that deuterium is only produced in significant amounts
during primordial nucleosynthesis and thoroughly destroyed in stellar interiors.
It is thus a key element in cosmology and for galactic chemical evolution. 
Amongst the different methods to measure the interstellar abundance of 
deuterium, the safest one is to observe simultaneously the atomic transitions
of D and H of the Lyman series in the far-UV, in absorption against the 
background continuum of stars (for a review, see e.g. Ferlet and Lemoine 1997).
In order to resolve the \is~velocity structure, we have already noted the need 
for the highest possible spectral resolution, such as the one offered by HST. 
Nevertheless, HST can only observe at \lya~for which \di~can only be detected 
(at $-$82 \kms~from \hi) for \hi~column densities smaller than 10$^{19}$ 
cm$^{-2}$, i.e. in the local \is~medium. Thus, only cool stars and white dwarfs
are available. Although the chromospheric emission line of cool stars has to be 
modeled, the Capella sight-line has provided the most precise measurement of 
the LIC D/H ratio: 1.60$\pm$0.09$^{+0.05}_{-0.10}\times$10$^{-5}$~(Linsky et 
al. 1995). However, the trivial velocity distribution of the Capella line of 
sight (a single cloud) has been derived through \feii~and \mgii~lines which may 
not properly trace the \hi~gas.

This is why we have chosen hot white dwarfs so as to provide a smooth stellar 
profile at \lya~and to also observe \ni~and \oi~lines which allow an accurate
kinematic sampling of the sight-line. The LIC has been clearly detected at the 
expected velocity towards G191--B2B, only 8$^o$~from Capella on the sky (Lemoine
et al. 1995), together with two other components. The sophisticated analysis of 
Vidal--Madjar et al. (1998) first confirms the Linsky et al. (1995) D/H value 
for the LIC and second concludes that D/H is of the order of 
0.9$\pm$0.1$\times$10$^{-5}$~on average for the two other components. Therefore,
the D/H ratio varies by nearly a factor of two over a few parsecs, raising thus
many striking questions beyond the scope of the present review.

Both the gas temperature and the Doppler parameter (b-value) of the LIC derived 
by Vidal--Madjar et al. (1998) are compatible with those quoted by Linsky et al.
(1995), although the most probable value for T is somewhat smaller. It has to
be noted that the commonly used procedure in \is~absorption studies is to fit 
Voigt profiles, accounting for radiative damping and Doppler broadening in the 
microturbulent limit (completely uncorrelated bulk motions). But it has been
suggested that a mesoturbulent model in which the influence of a finite 
correlation length in the stochastic velocity field on the line forming process
is accounted for might be more appropriate (e.g. Levshakov et al. 1997). 
Interpretation of observed profiles may then be substantially changed, with
important consequences in particular also with respect to deuterium abundances
towards quasars. The excellent agreement between the observed Doppler shifts
and those expected from projection of the LIC velocity vector can perhaps be
considered as the only proof that the usual microturbulent limit is not wrong.

It is worthwhile to emphasize that in the course of their G191--B2B study,
Vidal--Madjar et al. (1998) found for the two other non LIC components very 
different behaviors, namely very low and high temperatures ($\sim$2000 and
$\sim$11000 K) associated with respectively very high and low b-values 
($\sim$3 and $\sim$0.5 \kms). This could very well be the signature of an 
\is~shock, perhaps related to the other evidence already noted previously. 
\b\s

5 Hot hydrogen... a hydrogen wall?
\b

The physical parameters of the G cloud identified towards $\alpha$~Cen are
derived through profile fitting of the \feii, \mgii~and \di~lines in both
components of this visual binary system. However, the single-component fits to 
the \hi~lines yield for \hi~i) a large b-value which implies a temperature
significantly higher than for the other lines, and ii) a velocity redshift by
about 2.2 \kms~with respect to all the other lines, well above the HST--GHRS
accuracy (Linsky and Wood 1996). The most sensible way to resolve this 
discrepancy was to add a second absorption component to the \hi~lines, which 
has a temperature T=29000$\pm$5000 K, a smaller velocity relative to the LIC 
inflow and a column density log N(\hi)=14.74$\pm$0.24 \cm~much too low to be 
detected in the other lines. Linsky and Wood (1996) proposed that this gas is 
located near the heliopause and is the heated neutral hydrogen from the 
\ism~that is compressed by the solar wind (Fig. 1).

First shown theoretically by Baranov et al. (1991), the pile-up region of 
compressed \hi~in the upwind direction now called the "hydrogen wall" is 
predicted by multifluid gasdynamical models of the \is~gas and solar wind 
interaction, with physical parameters within the range of the above observed
parameters (e.g. Baranov and Malama 1995; Williams et al 1997). One can mention
that Qu\'emerais et al. (1995) found an increase in the diffuse \lya~emission 
in the upwind direction in their analysis of low dispersion spectra from the 
Voyager 1 and 2 spacecrafts, possibly due to the solar hydrogen wall. The 
Voyagers were then at 54 and 40 AU from the Sun.

Dring et al. (1997) and Wood and Linsky (1998) have tentatively detected a hot 
hydrogen component with a blueshifted velocity relative to the LIC inflow
towards five more nearby late-type stars that they interpret as produced by 
collisions between the stellar winds (instead of the solar wind) and the LIC 
flow. Within that framework, Wood and Linsky (1998) provide empirical 
measurements of wind properties for late-type main sequence stars. One should 
mention nevertheless that these authors have also found equally good fits of 
the \lya~lines without hot hydrogen absorption, but which require then 
characteristics either stellar (like profiles with a deep self-reversal) or 
\is~(like very low D/H ratios) that they judged improbable. However, in the 
previous section, we have seen that very low D/H values indeed seem to exist. 
Furthermore, such a hot \hi~absorption has already been found towards Sirius A,
with a slightly smaller velocity relative to the LIC inflow, which has been 
interpreted as due to an evaporation interface around the LIC (Bertin et al. 
1995). It has to be noted that Bertin et al. (1995) failed to detect \civ~as 
predicted by general conductive boundary models (Slavin 1989). But these models
also predict an emission spectrum which can apparently be rejected at the 
99.7\% confidence level by EUVE observations (Jelinsky et al. 1995). However, 
the inclusion of an \is~magnetic field would bring these models into agreement 
with EUVE data. Although a solar wind compression is attractive, an evaporation
interface is not precluded yet.

Revisiting the Capella line of sight and having confirmed the Linsky et al. 
(1995) results for the LIC, Vidal--Madjar et al. (1998) discovered an additional
weak and hot \hi~component which is also entirely compatible with the G191--B2B
fits. Again, it can be related either to a cloud interface or a "hydrogen wall".
In this latter case, it would be due to the Capella wind, implying thus a LIC 
extent all the way to Capella at 12.5 pc. This is not in contradiction
with previous sections if other kinematic components seen towards G191--B2B are 
either cloudlets embedded in the LIC or are interpreted in terms of a passing 
shock front.

\b\b
CONCLUSION
\b\s

The local \ism~kinematics indicates outflowing gas from the nearby ($\approx$170
pc) Scorpius--Centaurus Association. Over the past 15 million years, this OB 
Association has undergone three epochs of star formation. It has been shown that
Loop I is well described as a superbubble produced by the collective stellar 
winds and about 40 consecutive supernova explosions from this association (Egger
1995). Enhanced abundances of refractory elements further indicate that the
local gas has been processed through a shock front. Spectral signatures of such
a shock have been pointed out towards G191--B2B. Is this shock associated with 
the above superbubble? If it is clear that the Local Bubble should be no longer 
regarded as an isolated phenomenon, there is not even a rough agreement on its 
origin. Loop I and the Local Bubble can be viewed as due to two independent
supernova events that are now colliding, or the Local Bubble was sculpted by the
Sco--Cen activity which has swept up molecular clouds (see e.g. the discussion
and references in Breitschwerdt et al. 1996).

It is interesting to note that using fossil records from antarctic ice cores, 
Sonett et al. (1987) concluded that the observed $^{10}$Be concentrations, 
which show two "spikes" at about 35000 and 60000 years ago, gave strong 
evidence for a recent ($\leq$10$^5$~yrs) supernova explosion within a few tens 
of parsec from the Sun. If this is correct, a supernova is still expanding 
through the LIC.

The solar wind expands from the Sun at supersonic velocities. Since it depends 
on the solar latitude, it creates a non spherical cavity, the heliosphere, out 
to the termination shock. There is a transition region between the termination 
shock and a forward shock propagating into the \ism~which marks the changeover
to the LIC region. The external frontier of the heliosphere, the heliopause,
is at the pressure equilibrium between the two media. Taking a solar wind 
velocity of about 400 \kms~gives a stagnation point with respect to the LIC
pressure at about 500 AU. However, the estimated size of the heliosphere (based 
e.g. on indicators such as cosmic-ray modulation, anomalous component, upstream 
plasma waves) is substantially less, only of the order of 100 AU (as indicated 
in Fig. 1). Amongst several possibilities to resolve that discrepancy, a popular
one is pressure confinement through magnetic fields. A field strength of several
$\mu$G would be required, whereas the estimated local field is only 
$\sim$1.4$\mu$G (Frisch 1995). Conditions for general equilibrium seem not 
satisfied.

The location of the heliopause as well as the penetration of \is~matter into 
the solar system critically depend on the \is~magnetic field strength and the 
LIC ionization. Shielding the Earth, the heliospheric frontier is moving 
according to the local \is~environment changes. We can hope that the Voyager 
spacecrafts will encounter this frontier before vanishing around 2020.

\b\b
ACKNOWLEDGEMENT
\b\s

I am pleased to thank A. Vidal-Madjar for careful reading of the manuscript 
and J.P. Zimmermann for the drawings.

\b\b
REFERENCES
\bigskip
\s
\def\r{\par\noindent\hangafter=1\baselineskip=0.45cm\hangindent=1.2cm}

\r Baranov V., Lebedev M., Malama Y.: 1991, Astrophys. J., {\bf 375}, 347
\r Baranov V., Malama Y.: 1995, J. Geophys. Res., {\bf 100}, A8, 14755
\r Barstow M., Dobbie P., Holberg J., Hubeny I., Lanz T.: 1997, Month. Not. 
RAS, {\bf 286}, 58
\r Belfort P., Crovisier J.: 1984, Astron. Astrophys., {\bf 136}, 368
\r Bertaux J.L., Lallement R., Kurt V., Mironova E.: 1985, Astron. 
Astrophys., {\bf 150}, 82
\r Bertin P., Lallement R., Ferlet R., Vidal--Madjar A.: 1993a, J. Geophys. 
Res., {\bf 98}, A9, 15193
\r Bertin P., Lallement R., Ferlet R., Vidal--Madjar A.: 1993b, Astron. 
Astrophys., {\bf 278}, 549
\r Bertin P.: 1994, PhD Thesis, University of Paris VI
\r Bertin P., Vidal--Madjar A., Lallement R., Ferlet R., Lemoine M.: 1995, 
Astron. Astrophys., {\bf 302}, 889
\r Bowyer S., Lieu R., Sidher S., Lampton M., Knude J.: 1995, Nature, {\bf 375},
212
\r Breitschwerdt D., Egger R., Freyberg M., Frisch P., Vallerga J.: 1996, Sp.
Sci. Rev., {\bf 78}, 183
\r Breitschwerdt D., Freyberg M., Tr\"umper J. eds.: 1998, {\it The Local Bubble
and Beyond}, Lecture Notes in Physics, {\bf 506}, Springer
\r Cassinelli J., Cohen D., Macfarlane J., Drew J., Lynas--Gray A., Hubeny I.,
Vallerga J., Welsh B., Hoare M.: 1996, Astrophys. J., {\bf 460}, 949
\r Centurion M., Vladilo G.: 1991, Astrophys. J., {\bf 372}, 494
\r Cox D.: 1995, Nature, {\bf 375}, 185
\r Cox D., Reynolds R.: 1987, Ann. Rev. Astron. Astrophys., {\bf 25}, 303
\r Crawford I.: 1991, Astron. Astrophys., {\bf 247}, 183
\r Crawford I., Dunkin S.: 1995, Month. Not. RAS, {\bf 273}, 219
\r Crutcher R.: 1982, Astrophys. J., {\bf 254}, 82
\r Dame T., Ungerechts H., Cohen R., de Geus E., Grenier I., May J., Murphy D.,
Nyman L.A., Thaddeus P.: 1987, Astrophys. J., {\bf 322}, 706
\r Dring A., Murthy J., Henry R., Landsman W., Audouze J., Linsky J., Moos W.,
Vidal--Madjar A.: 1997, Astrophys. J., {\bf 488}, 760
\r Dupuis J., Vennes S., Bowyer S., Pradhan A., Thejll P.: 1995, Astrophys. 
J., {\bf 455}, 574
\r Egger R.: 1995, {\it The Physics of the Interstellar and the Intergalactic 
Medium}, eds. A. Ferrara, C. Heiles, C. McKee \& P. Shapiro, Astron. Soc. 
Pacific Conf. Series, {\bf 80}, 45
\r Ferlet R., Lallement R., Vidal--Madjar A.: 1986, Astron. Astrophys.,
{\bf 163}, 204
\r Ferlet R., Lecavelier des Etangs A., Vidal--Madjar A., Bertin P., Deleuil M.,
Lagrange--Henri A.M., Lallement R.: 1995, Astron. Astrophys., {\bf 297}, L5
\r Ferlet R., Lemoine M.: 1997, {\it Cosmic Abundances}, eds. S. Holt \& G. 
Sonneborn, Astron. Soc. Pacific Conf. Series, {\bf 99}, 78
\r Frail D., Weisberg J., Cordes J., Mathers C.: 1994, Astrophys. J., {\bf 436},
144
\r Franco G.: 1990, Astron. Astrophys., {\bf 227}, 499
\r Frisch P., York D.: 1983, Astrophys. J., {\bf 271}, L59
\r Frisch P., York D., Fowler J.: 1987, Astrophys. J., {\bf 320}, 842
\r Frisch P., York D.: 1991, {\it Extreme Ultraviolet Astronomy}, eds R. Malina
\& S. Bowyer, Pergamon Press, p. 322
\r Frisch P.: 1995, Sp. Sci. Rev., {\bf 72}, 499
\r Fruscione A., Hawkins I., Jelinsky P., Wiecigroch A.: 1994, Astrophys. J.
Suppl., {\bf 94}, 127
\r Genova R., Molaro P., Vladilo G., Beckman J.: 1990, Astrophys. J., {\bf 355},
150
\r Gry C., York D., Vidal--Madjar A.: 1985, Astrophys. J., {\bf 296}, 593
\r Gry C., Lemonon L., Vidal--Madjar A., Lemoine M., Ferlet R.: 1995, Astron.
Astrophys., {\bf 302}, 497
\r H\'ebrard G. et al.: 1998, in preparation
\r Hobbs L., Blitz L., Magnani L.: 1986, Astrophys. J., {\bf 306}, L109
\r Hunter S. et al.: 1997, Astrophys. J., {\bf 481}, 205
\r Jelinsky P., Vallerga J., Edelstein J.: 1995, Astrophys. J., {\bf 442}, 653
\r Jenkins E., Meloy D.: 1974, Astrophys. J., {\bf 193}, L121
\r Kondo Y., Bruhweiler F., Savage B. eds.: 1984, {\it Local Interstellar 
Medium}, IAU Coll. 81, NASA CP-2345
\r Lallement R., Vidal--Madjar A., Ferlet R.: 1986, Astron. Astrophys., 
{\bf 168}, 225
\r Lallement R., Ferlet R., Vidal--Madjar A., Gry C.: 1990, {\it Physics of the
Outer Heliosphere}, Cospar colloquia series, {\bf 1}, Pergamon Press, p. 37
\r Lallement R., Bertin P.: 1992, Astron. Astrophys., {\bf 266}, 479
\r Lallement R., Bertin P., Ferlet R., Vidal--Madjar A., Bertaux J.L.: 1994,
Astron. Astrophys., {\bf 286}, 898
\r Lallement R., Ferlet R., Lagrange A.M., Lemoine M., Vidal--Madjar A.: 1995, 
Astron. Astrophys., {\bf 304}, 461
\r Lallement R., Ferlet R.: 1997, Astron. Astrophys., {\bf 324}, 1105
\r Lemoine M., Vidal--Madjar A., Ferlet R., Bertin P., Gry C., Lallement R.:
1995, Astron. Astrophys., {\bf 308}, 601
\r Levshakov S., Kegel W., Mazets I.: 1997, Month. Not. RAS, {\bf 288}, 802
\r Linsky J., Diplas A., Wood B., Brown A., Ayres T., Savage B.: 1995, 
Astrophys. J., {\bf 451}, 335
\r Linsky J., Wood B.: 1996, Astrophys. J., {\bf 463}, 254
\r Lucke P.: 1978, Astron. Astrophys., {\bf 64}, 367
\r Magnani L., Blitz L., Mundy L.: 1985, Astrophys. J., {\bf 295}, 402
\r McCammon D., Sanders W.: 1990, Ann. Rev. Astron. Astrophys., {\bf 28}, 657
\r McKee C., Ostriker J.: 1977, Astrophys. J., {\bf 218}, 148
\r Mihalas D., Binney J.: 1981, {\it Galactic Astronomy, Structure and 
Kinematics}, 2$^{nd}$~Edition, Freeman \& Co.
\r Osterbart R., Fahr H.: 1992, Astron. Astrophys., {\bf 264}, 260
\r Paresce F.: 1984, Astron. J., {\bf 89}, 1022
\r Piskunov N., Wood B., Linsky J., Dempsey R., Ayres T.: 1997, Astrophys. J., 
{\bf 474}, 315
\r Puyoo O., Ben Jaffel L., Emerich C.: 1997, Astrophys. J., {\bf 480}, 262
\r Qu\'emerais E., Lallement R., Bertaux J.L.: 1992, Astron. Astrophys., 
{\bf 265}, 806
\r Qu\'emerais E., Bertaux J.L., Sandel B., Lallement R.: 1994, Astron. 
Astrophys., {\bf 290}, 941
\r Qu\'emerais E., Sandel B., Lallement R., Bertaux J.L.: 1995, Astron.
Astrophys., {\bf 299}, 249
\r Reynolds R.: 1990, Astrophys. J., {\bf 348}, 153
\r Routly P., Spitzer L.: 1952, Astrophys. J., {\bf 115}, 227
\r Sanders W., Edgar R.,  Liedahl D.: 1996, {\it R\"ontgenstrahlung from the 
Universe}, MPE Report {\bf 263}, 339
\r Slavin J.: 1989, Astrophys. J., {\bf 346}, 718
\r Snowden S., Cox D., McCammon D., Sanders W.: 1990, Astrophys. J., {\bf 354},
211
\r Snowden S., McCammon D., Verter F.: 1993, Astrophys. J., {\bf 409}, L21
\r Snowden S., Mebold U., Herbstmeier U., Hirth W., Schmitt J.: 1991, Science, 
{\bf 252}, 1529
\r Sonett C., Morfill G., Jokipii J.: 1987, Nature, {\bf 330}, 458
\r Spitzer L.: 1985, Physica Scripta, {\bf T11}, 5
\r Stacy J., Jackson P.: 1982, Astron. Astrophys. Suppl., {\bf 50}, 377
\r Tinbergen J.: 1982, Astron. Astrophys., {\bf 105}, 53
\r Vallerga J., Vedder P., Craig N., Welsh B.: 1993a, Astrophys. J., {\bf 411},
729
\r Vallerga J., Vedder P., Welsh B.: 1993b, Astrophys. J., {\bf 414}, L165
\r Vallerga J.: 1996, Sp. Sci. Rev., {\bf 78}, 277
\r Vidal--Madjar A., Laurent C., Bruston P., Audouze J.: 1978, Astrophys. J.,
{\bf 223}, 589
\r Vidal--Madjar A., Lemoine M., Ferlet R., H\'ebrard G., Koester D., Audouze 
J., Cass\'e M., Vangioni--Flam E., Webb J.: 1998, Astron. Astrophys., {\bf
338}, 694
\r Welsh B., Vedder P., Vallerga J.: 1990, Astrophys. J., {\bf 358}, 473
\r Welsh B., Vedder P., Vallerga J., Craig N.: 1991, Astrophys. J., {\bf 381},
462
\r Welty D., Morton D., Hobbs L.: 1996, Astrophys. J. Suppl., {\bf 106}, 533
\r Welty D.: 1998, {\it The Local Bubble and Beyond}, eds D. Breitschwerdt, M.
Freyberg \& J. Tr\"umper, Lecture Notes in Physics, {\bf 506}, 151
\r Williams L., Hall D., Pauls H., Zank G.: 1997, Astrophys. J., {\bf 476}, 366
\r Witte M., Rosenbauer H., Banaszkiewicz M.: 1993, Adv. Space Res. 6, {\bf 13},
121
\r Wood B., Linsky J.: 1997, Astrophys. J., {\bf 474}, L39
\r Wood B., Linsky J.: 1998, Astrophys. J., {\bf 492}, 788

\b\b
FIGURES CAPTIONS
\b\s

{\bf Figure 1}: Schematic illustration of the solar wind--interstellar medium
interface. The Sun is moving through the LIC and the supersonic solar wind is
blowing a cavity called the heliosphere. The fraction of the interstellar 
neutral hydrogen which penetrates within the heliosphere is decelarated through 
charge--exchange collisions, invades the interplanetary medium and backscatters 
the solar \lya~radiation producing an observed emission glow. The interstellar
neutral helium is extremely weakly interacting with the plasma and keeps its
initial heliocentric velocity, measured {\it in situ} by Ulysses and found to be
equal to the LIC velocity. The transition from supersonic to subsonic flow is
supposed to occur at the so-called termination shock, the first discontinuity
the outer probes Voyager and Pioneer 11 should encounter.

\b

{\bf Figure 2}: Schematic morphology of the very Local Interstellar Medium. The
Sun is embedded in the LIC (assumed to be a cloudlet with a more or less 
constant density), very close to its boundary in the direction of $\alpha$~Cen. 
This line of sight exhibits a single absorption component at a non-LIC velocity.
Cloud multiplicity towards other nearby stars is illustrated. Some kinematical 
information is also given (in the LSR). It is consistent with an expanding 
medium from the direction of the Galactic Center, however the 3D velocity 
vectors are known only for the LIC and the G cloud. The exact location and 
shape of the different cloudlets are unknown, but filamentary structures are 
not unusual in interstellar gas. These entities are embedded in the very low 
density hot Local Bubble, which extends in particular beyond Sirius towards 
CMa.

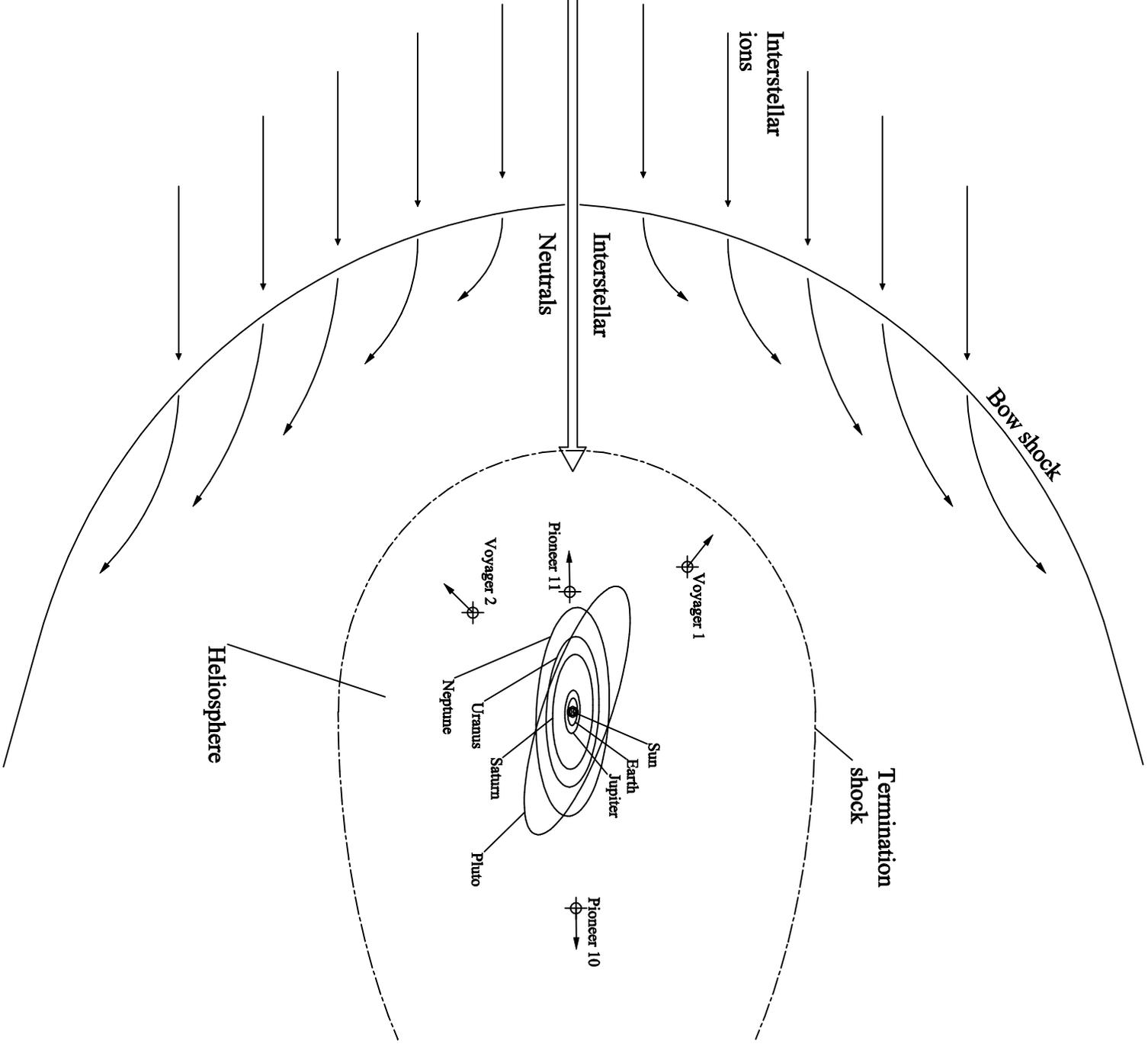
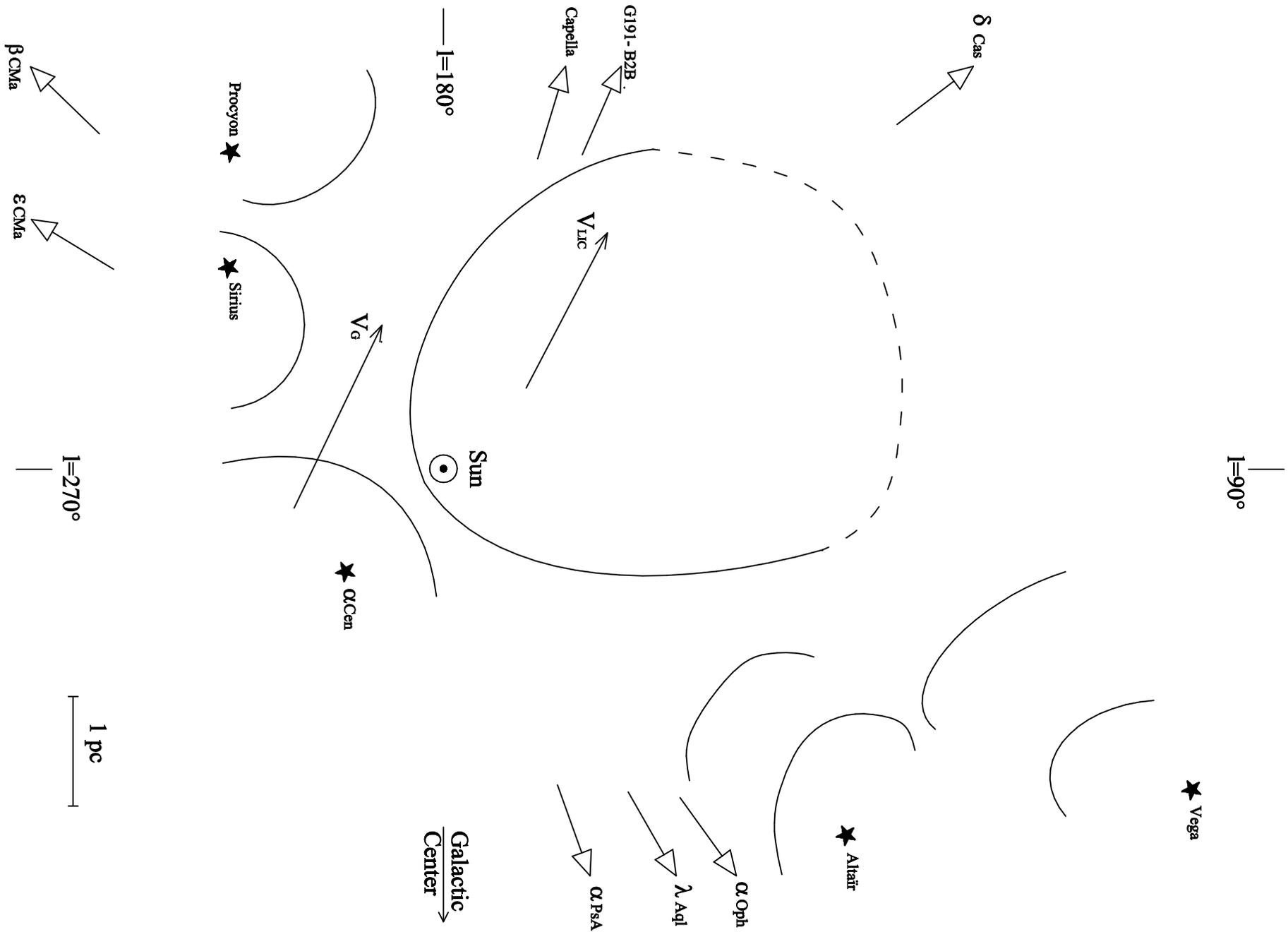
\end{document}